\documentclass[%
 reprint,
 amsmath,amssymb,
 aps,
]{revtex4-2}

\usepackage{graphicx}
\usepackage{graphicx, xurl, hyperref}
\usepackage{color}
\usepackage{dcolumn}
\usepackage{bm}

\usepackage{titlesec}

\titlespacing*{\section}{0pt}{10pt}{10pt}  
\titlespacing*{\subsection}{0pt}{8pt}{8pt}  
\titlespacing*{\subsubsection}{0pt}{6pt}{6pt} 

\begin{document}

\preprint{APS/123-QED}

\title{A machine-learning study of phase transitions in Ising, Blume-Capel, and Ising-metamagnet models}

\author{Vasanth Kumar Babu}
	\email[]{vasanthb@iisc.ac.in}
		\affiliation{Centre for Condensed Matter Theory, Department of Physics, Indian Institute of Science, Bangalore, 560012, India. }
	\author{Rahul Pandit}
		\email[]{rahul@iisc.ac.in}
	\affiliation{Centre for Condensed Matter Theory, Department of Physics, Indian Institute of Science, Bangalore, 560012, India. }
 
\date{\today}

\begin{abstract}
We combine  machine-learning (ML) techniques with Monte Carlo (MC) simulations and finite-size scaling (FSS) to study continuous and first-order phase transitions in Ising, Blume-Capel, and Ising-metamagnet spin models. We go beyond earlier studies that had concentrated on obtaining the correlation-length exponent $\nu$. In particular,  we show (a) how to combine neural networks (NNs), trained with data from MC simulations of Ising-type spin models on finite lattices, with FSS to obtain both thermal magnetic exponents $y_t = 1/\nu$ and $y_h$, respectively, at both critical and tricritical points, (b) how to obtain the NN counterpart of two-scale-factor universality
at an Ising-type critical point, and (c) FSS at a first-order transition. We also obtain the FSS forms for the output of our trained NNs as functions of both the temperature and the magnetic field.
\end{abstract}

\maketitle


\section{\label{sec:Introduction}Introduction}

The development of the theory of phase transitions is among the most important advances in theoretical physics over the last sixty years~\cite{fisher1974renormalization,fisher1983scaling,kadanoff1993critical,fisher1998renormalization,domb2000phase,kadanoff2000statistical,kardar2007statistical,goldenfeld2018lectures}. We can now obtain, with great accuracy, the universal critical exponents that characterize the universality classes of different continuous phase transitions. Examples of such transitions include the liquid-gas critical point, the Curie and N\'eel transitions in ferromagnets and antiferromagnets, respectively, the normal-superconductor transition, and the change from a normal fluid to a superfluid.  

Some studies have explored intriguing intersections between statistical physics and machine learning (ML). Ideas from statistical physics are frequently employed to gain insights into ML models, even as the application of ML methods in physics moves apace  [see, e.g., Refs.~\cite{mezard2009information,carleo2019machine,mehta2019high}]. A diverse array of ML techniques are being utilized to explore properties of various models of statistical physics, such as Ising models ~\cite{carrasquilla2017machine,morningstar2018deep, li2019extracting,zhang2019machine,kim2021emergence,shiina2020machine,chertenkov2023finite,bayo2025machine}, directed percolation~\cite{zhang2019machine,shen2021supervised,bayo2022machine,bayo2023percolating,wang2024supervised}, and models for nonequilibrium transitions~\cite{venderley2018machine,tang2024learning}. Furthermore, the ML-aided classification of phases and phase transition has attracted significant attention~\cite{ tanaka2017detection,hu2017discovering,miyajima2021machine, suchsland2018parameter,zhang2019few,li2019extracting,zhang2019machine,tian2023machine,carrasquilla2017machine,kim2021emergence,shen2021supervised,shiina2020machine,chertenkov2023finite,wang2024supervised}; both supervised and unsupervised ML methods have been used in such classification. For example, ML-assisted dimensionality reduction, in the vicinity of critical points, has used principal component analysis (PCA) and autoencoders  [see, e.g., ~\cite{wang2016discovering,hu2017discovering,mendes2021unsupervised}]. 

Phase transitions can occur only in the thermodynamic limit, i.e., when the linear system size $L \to \infty$. However, finite-size-scaling (FSS) analysis provides an effective method for estimating, from finite-size calculations, thermodynamic functions and their singularities, for both continuous and first-order transitions~\cite{fisher1972scaling, barber1983finite, goldenfeld2018lectures,fisher1982scaling,binder1984finite}. Traditionally, FSS is used directly with thermodynamic functions like the magnetization $M$ in a ferromagnet. Recent studies have demonstrated that outputs from machine-learning models, such as neural networks (NNs), can be analysed with FSS to obtain thermal critical exponents, like $\nu$, at Ising-type critical points~\cite{carrasquilla2017machine,  li2019extracting,kim2021emergence,shen2021supervised,shiina2020machine,chertenkov2023finite,wang2024supervised}. We extend these studies significantly by working with the two-dimensional Ising, Blume-Capel, and Ising-metamagnet models and demonstrating (a) how to combine NNs, trained with data from Monte-Carlo simulations of Ising-type spin models on finite lattices, with FSS to obtain both thermal magnetic exponents $y_t = 1/\nu$ and $y_h$, respectively, at \textit{both critical and tricritical points}, (b) how to obtain the NN counterpart of two-scale-factor universality~\cite{PhysRevLett.29.345,PhysRevB.9.2107} at an Ising-type critical point, and (c) how to combine NN methods and FSS at a first-order transition. We also use the probability distribution functions (PDFs) of the magnetization $M$, near the critical point, to obtain general forms for the output of our trained NNs as functions of the temperature and the magnetic field; this has not been attempted hitherto.

\vspace{0.3cm}
 The remaining part of our paper is organized as follows: In Sec.~\ref{sec:methods} we define the Ising, Blume-Capel, and Ising-metamagnet models, the Monte-Carlo and finite-size scaling methods we use, and the neural network architectures we employ. We then present our results in Sec.~\ref{sec:Results}. Section~\ref{sec:conclusions} is devoted to
 a discussion of our results and conclusions.
 
\section{Models and Methods\label{sec:methods}}

\subsection{Models}

We consider the following three spin models for two-dimensional (2D) square lattices: the ferromagnetic Ising model [Subsection~\ref{subsec:Ising}], the Blume-Capel model [Subsection~\ref{subsec:BC}], and the Ising-metamagnet [Subsection~\ref{subsec:MM}]. 

\subsubsection{Ising ferromagnet}
\label{subsec:Ising}

The Ising model is defined by the Hamiltonian 
\begin{equation}
\mathcal{H}_I
= - J \sum_{\langle i,j \rangle} S_{i} S_{j} - H \sum_{i} S_{i}\,,
\label{eq:Isingham}
\end{equation}
where the Ising spins $S_i = \pm 1$ and $\langle i,j \rangle$ denotes nearest-neighbor 
pairs of sites~\cite{onsager1944crystal,mccoy1973two,baxter2016exactly,thompson2015mathematical}
and we consider a square lattice with $N=L^2$ sites, labeled by $i$. There is a magnetic field $H$ at every site; and the exchange coupling $J > 0$, i.e., we consider the ferromagnetic case, which has completely aligned spins at temperature $T=0$ and, at $H=0$, two coexisting phases: the $\uparrow$ phase with $S_i = 1,\,\, \forall i$; and the $\downarrow$ phase with $S_i = -1,\,\, \forall i$. The equilibrium statistical mechanics of this model follows from its intensive bulk free energy $f_B$, which can be obtained exactly~\cite{onsager1944crystal,mccoy1973two,baxter2016exactly,thompson2015mathematical} if $H=0$, whence we know that this Ising model exhibits a critical point at $H=0$ and $T=T_c^{eq}$, with 
\begin{eqnarray}
 \sinh \bigg[\frac{2J}{k_BT_c^{eq}}\bigg] &=& 1\,, \nonumber \\
 \implies k_BT_c^{eq} &=& \frac{2J}{\ln(1+\sqrt{2})}\,,
 \label{eq:IsingTc}
\end{eqnarray}
where $k_B$ is the Boltzmann constant. Henceforth, we use units in which $k_B$ and $J$ are set to $1$. 
The \textit{order parameter} for this model is the magnetization per spin $M =\langle \sum_i S_i \rangle/N$, where the angular brackets denote thermal averages like $\langle S_i \rangle = \sum_{\{S_j\}} [S_i \exp(-\beta \mathcal{H}_I)]/\sum_{\{S_j\}} [\exp(-\beta \mathcal{H}_I)]$, where $\sum_{\{S_j\}}$ denotes the sum over all spin states, and $\beta \equiv 1/(k_BT)$. This magnetization and the two-spin correlation length $\xi$ show the following power-law behaviors in the vicinity of the critical point:
\begin{eqnarray}
M &\sim& |t|^{\beta} \,, \;\; t\to 0^- \text{and}\;\; h=0\,; \nonumber \\
M &\sim& |h|^{1/\delta} \,, \;\; h\to 0^{\pm} \text{and}\;\; t=0\,; \nonumber \\
\xi &\sim& |t|^{-\nu} \,, \;\; t\to 0^{\pm} \text{and}\;\; h=0\,; \nonumber \\
\xi &\sim& |h|^{-\beta\delta/\nu} \,, \;\; h\to 0^{\pm}\;\; \text{and}\;\; t=0\,;
\label{eq:IsingExponents}
\end{eqnarray}
the reduced temperature $t=\frac{T-T_{c}}{T_{c}}$; and $t\to 0^-$ indicates that $t$ approaches $0$ from below; $t\to 0^{\pm}$ means that $t$ approaches $0$ either from above or below. The exponents $\beta =1/8$, $\nu=1$, and $\delta=15$ for this 2D Ising model (and all models in this universality class~\cite{onsager1944crystal,mccoy1973two,baxter2016exactly,thompson2015mathematical,kardar2007statistical,goldenfeld2018lectures}).

\subsubsection{Blume-Capel model}
\label{subsec:BC}

We also study the Blume-Capel model on a 2D square lattice with nearest-neighbor interactions and the Hamiltonian~\cite{beale1986finite,wilding1996tricritical,kwak2015first,moueddene2024critical}
\begin{equation}
\mathcal{H}_{BC}
    =  - J \sum_{\langle i,j \rangle} \mathcal{S}_{i} \mathcal{S}_{j} + \Delta \sum_{i}  \mathcal{S}_{i}^{2} - H \sum_{i} \mathcal{S}_{i}  \,,
\label{eq:IsingBC}
\end{equation}
where the spins $\mathcal{S}_i = \pm 1, 0$, the ferromagnetic coupling $J > 0$, and $\Delta$ and $H$ are, respectively, the crystal field and the magnetic field $H$. In the $\Delta-T$ plane, with the $H=0$, this model exhibits a line of first-order transitions and another line of 2D Ising-type second-order transitions that meet at a tricritical point~\cite{beale1986finite,wilding1996tricritical,kwak2015first,moueddene2024critical}, which has distinct critical exponents. This model~\eqref{eq:IsingBC} has the following two order parameters:
\begin{equation}
M \equiv \frac {\langle \sum_i \mathcal{S}_i \rangle}{N}\,;\;\; Q \equiv \frac {\langle \sum_i \mathcal{S}_i^2 \rangle}{N}\,.
\label{eq:BCOPs}
\end{equation}

\subsubsection{Ising-metamagnet} 
\label{subsec:MM}

We consider the Ising-metamagnet on a two-dimensional (2D) square lattice,  with $N=L^2$ sites, labelled by $i$,  nearest-neighbor ($nn$) anti-ferromagnetic interactions  $J_{1}<0$, and next-nearest-neighbor ($nnn$) ferromagnetic interactions $J_{2} > 0$, and the Hamiltonian~\cite{PhysRevB.28.2686,beale1984finite,landau1981tricritical,herrmann1984finite}
\begin{eqnarray}
 \mathcal{H}_{M}
&=&  -J_{1} \sum_{{\langle i,j \rangle}_{nn}} S_{i} S_{j} - J_{2} \sum_{{\langle i,j \rangle}_{nnn}} S_{i} S_{j}\nonumber  \\
&-& H \sum_{i} S_{i}- H_{s}[ \sum_{i\in A} S_{i} - \sum_{i\in B} S_{i} ] \,,\label{eq:IsingMeta}
\end{eqnarray}
where the Ising spins $S_i = \pm 1$;  $H$ and $H_s$ are, respectively, the external magnetic field and the  \textit{staggered} magnetic field;  $A$ and $B$ are the two interpenetrating square sublattices that comprise our original bipartite square lattice. 

The fields $H$ and $H_s$ are thermodynamically conjugate to the magnetization $M$ and the \textit{staggered} magnetization $M_{s}$ order parameters for this system; these are defined as follows:
\begin{eqnarray}
M_A &\equiv& \frac { \sum_{i\in A} \langle S_i \rangle}{N_A}\,;\;\;M_B \equiv \frac {\sum_{i\in B} \langle S_i \rangle}{N_B}\,;\nonumber \\
M &\equiv& M_A+M_B \,;\;\; M_s \equiv M_A-M_B\,;
\label{eq:MMOP}
\end{eqnarray}
here, $N_A = N_B = N/2$.

In the $H-T$ plane, with $H_{s}=0$ and $H \geq 0$, the model~\eqref{eq:IsingMeta} exhibits a line of first-order transition and a line of  second-order transitions, in the 2D Ising universality class, that meet at a tricritical point~\cite{PhysRevB.28.2686,beale1984finite,landau1981tricritical,herrmann1984finite}.

\subsection{Monte Carlo Simulations}

For arbitrary values of the exchange couplings $J, \ldots$, fields $H, \dots$, and the temperature $T$, the bulk free energy $f_B$ and other thermodynamic functions like $M$  cannot be obtained analytically for the models~\eqref{eq:Isingham}, \eqref{eq:IsingBC}, and \eqref{eq:IsingMeta}. Here, we use standard Metropolis Monte Carlo methods~\cite{landau2021guide}, with non-conserved order parameters, to obtain order parameters and their probability distributions. Furthermore, we use various spin configurations, which emerge from our MC simulations at given values of couplings, fields, and the temperature $T$, as inputs for training and testing the convolutional neural networks (CNNs) and fully connected neural networks (FCNNs) [see Subsection~\ref{subsec:NN}.   

\subsection{Finite-size scaling}
\label{subsec:FSS}

Finite-size scaling helps us to extract universal critical exponents and scaling functions, in the vicinity of a critical or a tricritical point, from calculations on small systems, e.g., with $N$ small in models~\eqref{eq:Isingham}, \eqref{eq:IsingBC}, and \eqref{eq:IsingMeta}. Strictly speaking, to obtain the intensive bulk free energy $f_B$, we must take the thermodynamic limit $L\to\infty$, where the linear system size $L = Na$, with $a$ the lattice spacing for the models we consider (we choose $a=1$). In the vicinity of the critical point, $f_L$, the finite-size approximation to the singular part of $f_B$, assumes a scaling form~\cite{barber1983finite,privman1990finite}, which is given below for model~\ref{eq:Isingham}: 
\begin{equation} \label{eq:fss}
f_{L}(t,h)  = L^{-d}f_{s}(tL^{y_{t}},hL^{y_{h}})\,,
\end{equation}
where $d$ is the dimension, the reduced temperature $t\equiv\frac{(T - T_{c})}{T_{c}}$, with $T_{c}$ the critical temperature, $h\equiv H/(k_BT_c)$, and $y_{t}$ and $y_{h}$ are universal scaling exponents that are related to the conventional critical exponents via
\begin{eqnarray}
y_{t}&=&\frac{1}{\nu}\;\; {\rm{and}}\;\; y_{h}=\beta\delta/\nu \;\; {\rm{whence}}\;\; \label{scalrel}\\
M &=& L^{-\beta/\nu} f_{s}(tL^{1/\nu})\,,\;\; {\rm{at}} \;\; h = 0\,; \label{eq:fssmt} \\
M &=& L^{-\beta/\nu} f_{s}(hL^{\beta\delta/\nu})\,,\;\; {\rm{at}} \;\; t = 0\,. \label{eq:fssmh}
\end{eqnarray}
From Eqs.~\eqref{eq:fssmt} and ~\eqref{eq:fssmh}, we can see that plots of (a) $ML^{\beta/\nu}$ vs $tL^{1/\nu}$, for $h=0$,  and (b) $ML^{\beta/\nu}$ vs $hL^{\beta\delta/\nu}$, for $t=0$, fall on top of scaling curves for different values of $L$ [these curves are different for cases (a) and (b)]. The best fits for these scaling curves lead to estimates for the critical exponents.

For the 2D Blume-Capel  model~\ref{eq:IsingBC},  in the vicinity of  tricritical point~\cite{beale1986finite,kwak2015first,moueddene2024critical}at $H=0$, $T=T^{BC}_{t}$, and $\Delta=\Delta_t$, the finite-size scaling relation for the singular part of the free energy is
\begin{equation} 
f^{BC}_{L}(t,g,h)  = L^{-d}f_{L}(tL^{y_{t}},gL^{y_{g}},hL^{y_{h}})\,,
\label{eq:FBC_tri}
\end{equation}
where $t=\frac{(T-T^{BC}_{t})}{k_{B}T^{BC}_t}$ and $g=\frac{(\Delta-\Delta_t)}{(k_BT^{BC}_t)}+a t$~\cite{beale1986finite,moueddene2024critical}, where $a$ is a nonuniversal constant, and $g$ is the deviation from the tricritical point along the tangent to the coexistence curve, in the $t-\Delta$ plane.

Similarly, for the 2D metamagnet, the singular part of the free energy in the vicinity of the tricritical point, at  $H_s = 0$, $T=T^{M}_{t}$, and $H=H^{M}_t$,  has the form [cf. Refs.~\cite{beale1984finite,landau1981tricritical,herrmann1984finite}]
\vspace{-0.1cm}
\begin{equation} 
f^{M}_{L}(t,g',h_{s})  = L^{-d}f_{L}(tL^{y_{t}},g'L^{y_{g'}},h_{s}L^{y_{h}})\,,
\label{eq:FBC_tri}
\end{equation}
where $t=\frac{(T-T^{M}_{t})}{k_{B}T^{M}_t}$, and $g'=\frac{(H-H^{M}_t)}{(k_BT^{M}_t)}+a't$~\cite{herrmann1984finite} is the tangent to the co-existence curve in the $T-H$ plane, $h_{s} = H_s/(k_BT^{M}_t)$ is the staggered field; $a'$ is a nonuniversal coefficient.

\vspace{-0.1cm}
\subsection{Neural-network architectures}
\label{subsec:NN}
\begin{figure*}
    \centering  \includegraphics[width=\textwidth]{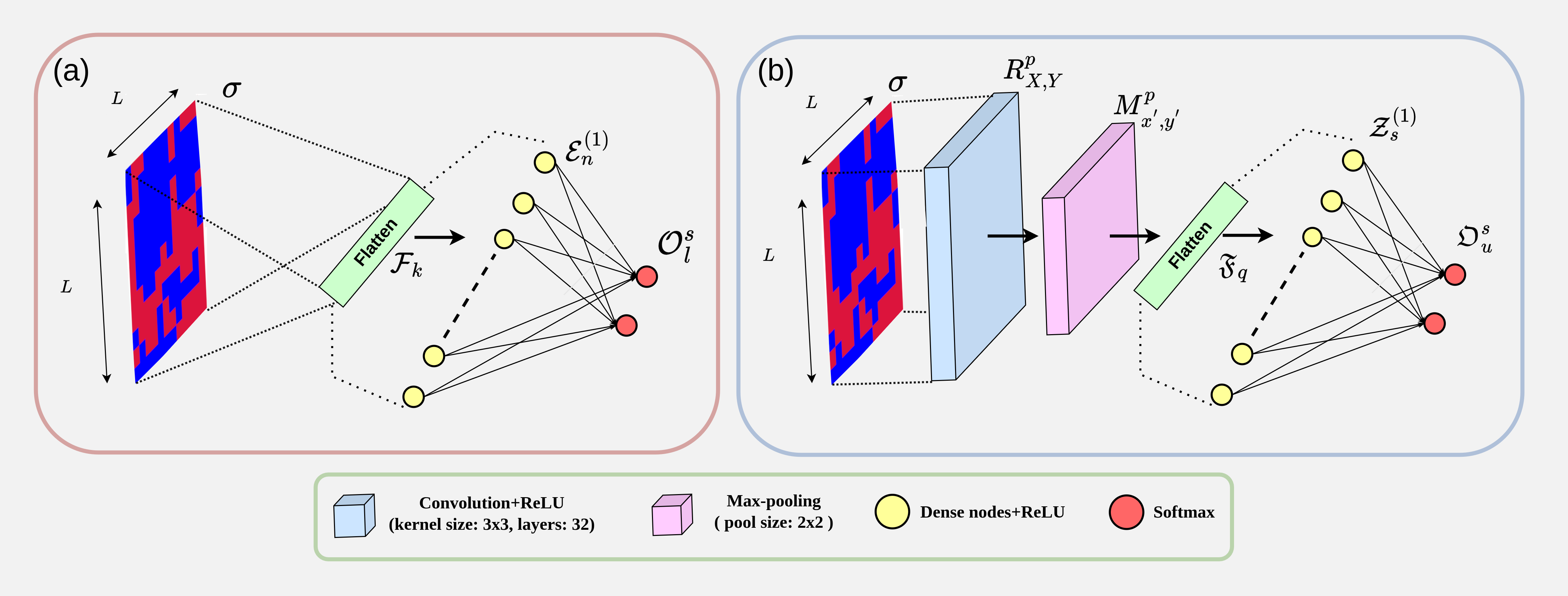}
    \caption{Schematic diagrams of (a) the fully connected neural nework (FCNN) and (b) the convolutional neural network (CNN), which we train to classify snapshots of spin configurations that we obtain from our Monte Carlo simulations, both above and below the critical temperature $T_c$. }
    \label{fig:CNNArch}
\end{figure*}

We employ fully connected neural networks (FCNNs) and convolutional neural networks (CNNs). In Fig.~\ref{fig:CNNArch} (a), we give a schematic diagram of the FCNN we use; the mathematical operations for  our FCNN are given below:
\begin{eqnarray}
\rm{Input}&:& \;\; \sigma_{X,Y} \;( L \times L )\,;\;\; {\rm{i.e.,}}\; 1 \leq X,Y \leq L;\nonumber \\
\rm{Flatten}&:& \;\;  \mathcal{F}_{k}  = \sigma_{X,Y}\,;\;\; k=(Y-1)L+X; \nonumber\\
\rm{Dense\;layer}\, 1&:& \;\; E^{(1)}_{n} =  \sum\limits_{k=1}^{L^2} W_{n,k}^{(1)} \mathcal{F}_{k}+b^{(1)}_{n};1\leq n \leq64\,;\nonumber \\
\rm{ReLU}&:& \;\; \mathcal{E}^{(1)}_{n}=max(0,E^{(1)}_n) ; 1\leq n \leq 64\,;\nonumber \\
\rm{Dense\;layer}\, 2&:& \;\; E^{(2)}_{l} =  \sum\limits_{m=1}^{64} W_{l,m}^{(2)} \mathcal{E}^{(1)}_{m}+b^{(2)}_{l}; 1\leq l \leq 2\,;\nonumber \\
\rm{Softmax}&:& \;\; \mathcal{O}^{s}_{l} =\frac{\exp(E^{(2)}_{l})}{\sum\limits_{j=1}^{2}\exp(E^{(2)}_{j})};\; l=1,2 \,. 
\label{eq:FCNN}
\end{eqnarray}
Here, the input $\sigma_{X,Y}$ is the spin at site $(X,Y)$ [$S_i$ for models~\eqref{eq:Isingham} and \eqref{eq:IsingMeta} and $\mathcal{S}_i$ for model~\eqref{eq:IsingBC}]. We have a hidden layer with $64$ dense nodes with weights $W^{(1)}_{n,k}$, and biases $b^{(1)}_{n}$, followed by the ReLU activation.  In the output, we have  $2$ nodes with weights $W^{(2)}_{l,m}$ and biases $b^{(2)}_{l}$, followed by a softmax activation to obtain the normalized outputs $\mathcal{O}^{s}_{l}$.

In Fig.~\ref{fig:CNNArch} (b) we give a schematic diagram of the CNN we use; and the mathematical operations for our CNN are given below: 
\begin{eqnarray}
\rm{Input}&:& \;\; \sigma_{X,Y} \;( L \times L )\,;\;\; {\rm{i.e.,}}\; 0\leq X,Y < L;\nonumber \\ 
\rm{Padding}&:& \;\; {\tilde{\sigma}}_{X,Y}\;\,;\;\; {\rm{i.e.,}}\; 0\leq X, Y< L+2;\nonumber \\ 
&&\;\;{\tilde{\sigma}}_{X,Y} = \sigma_{X-1,Y-1},\,\text{if}\,1\leq X,\,Y \leq L,\,\nonumber\\
&&\;\;\text{else}\,{\tilde{\sigma}}_{\tilde{X},\tilde{Y}}=0
\nonumber\\
\rm{Convolution}&:& \;\; C^{(p)}_{X,Y} = \sum_{i=0}^{2}\sum_{j=0}^{2}\tilde{\sigma}_{X+i,Y+j} F_{i,j}^{(p)}+b^{(p)}\,;\nonumber\\
&& \;\;p \in [0,1,\ldots,31]\,;\nonumber\\
\rm{ReLU}&:& \;\;  R^p_{X,Y} =  max(0,C^{(p)}_{X,Y}); \nonumber \\
\rm{Max-pool}&:& \;\;  M^{(p)}_{x^{'},y^{'}} = max(R^{(p)}_{(X:X+1)(Y:Y+1)})\,;\nonumber\\
&&2x'=X,\; 2y'=Y\,;\;\;0 \leq x',y'<\frac{L}{2}\nonumber\,;\\
\rm{Flatten}&:& \;  \mathfrak{F}_{q} = M_{x',y'}^{(p)};\;q=p(\frac{L}{2})^{2}+y'(\frac{L}{2})+x'\,;\;\nonumber\\
\rm{Dense\;layer}\, 1&:& \;\; Z^{(1)}_{s} =  \sum \limits_{r=0}^{32({\frac{L}{2}})^{2}-1}W_{s,r}^{(1)}\mathfrak{F}_{r}+b_{s}^{(1)};\;1\leq s \leq 64\,;\nonumber \\
\rm{ReLU}&:& \;\; \mathcal{Z}^{(1)}_{s} =  max(0,Z^{(1)}_{s}) ; 1\leq s \leq 64\,;\nonumber\\
\rm{Dense\;layer}\, 2&:& \;\; Z^{(2)}_{t} =  \sum\limits_{s=1}^{64} W_{t,s}^{(2)}\mathcal{Z}^{(1)}_{s}+ b_{t}^{(2)};\; 1\leq t \leq 2\,;\nonumber \\
\rm{Softmax}&:& \;\; \mathfrak{O}^{s}_{u} =\frac{\exp(Z_{u}^{(2)})}{\sum\limits_{t=1}^{2}\exp(Z_{t}^{(2)})};\; u =1,2\,;
\label{eq:CNN}
\end{eqnarray}
We have a convolutional layer, with $32$, $3\times3$ filters $F^{(p)}_{i,j}$, and biases $b^{(p)}$, with ReLU activation [we apply the padding to retain the original spatial dimensions], followed by a max-pooling layer with $2\times2$ filters and a stride of $2$, and a dense layer with $64$ nodes, with weights $W_{s,r}^{(1)}$, and biases $b_{s}^{(1)}$. In the output, we have  $2$ nodes with weights $W^{(2)}_{t,s}$ and biases $b^{(2)}_{t}$, followed by a softmax activation to obtain the normalized outputs $\mathfrak{O}^{s}_{t}$.

With the spin configurations $\sigma_{X,Y}$ as the input, we train our neural networks to classify these configurations using binary cross-entropy, with a regularizer term as the loss function,
\begin{eqnarray}
\mathcal{L} &=& \left\langle-P(\boldsymbol{\sigma})\log\hat{P}(\boldsymbol{\sigma})-(1-P(\boldsymbol{\sigma}))\log(1-\hat{P}(\boldsymbol{\sigma}))\right\rangle\,\nonumber\\
&+& \lambda ||\mathcal{W}||\,,
\label{eq:loss}
\end{eqnarray}
where $P(\boldsymbol{\sigma})$ takes the values $1$ (or $0$) if the spin configuration $\boldsymbol{\sigma}$, from the training data, is below (or above) the critical temperature; and $\hat{P}(\boldsymbol{\sigma}) \in [0,1]$ is the output or the prediction of the neural network, which is $\mathcal{O}^{s}_{1}$ for our FCNN [Eq.~\eqref{eq:FCNN}] and $\mathfrak{O}^{s}_{1}$ for our CNN [Eq.~\eqref{eq:CNN}], $\langle\cdot\rangle$ denotes the average over the training data set, $\lambda$ is the regularization strength, and
\begin{eqnarray}
||\mathcal{W}||&=&\sum_{n=1}^{64}\sum_{k=1}^{L^{2}}{W^{(1)}_{n,k}}^{2}+\sum_{n=1}^{64}{b^{(1)}_{n}}^{2}+\sum_{l=1}^{2}\sum_{m=1}^{64}{W^{(2)}_{l,m}}^{2}\,\nonumber\\
&+&\sum_{l=1}^{2}{b^{(2)}_{l}}^{2}\,,
\label{L2reg:FCNN}
\end{eqnarray}
for the FCNN, and 
\begin{eqnarray}
||\mathcal{W}||&=&\sum_{p=0}^{p=31}\sum_{j=0}^{2}\sum_{i=0}^{2}{F^{(p)}_{i,j}}^{2} 
+\sum_{p=0}^{p=31}{b^{(p)}}^{2}+\sum_{s=1}^{64}\sum_{r=0}^{32{(\frac{L}{2}})^{2}-1}{W^{(1)}_{s,r}}^{2}\,\nonumber\\
&+&\sum_{s=1}^{64}{b^{(1)}_{s}}^{2}+\sum_{t=1}^{2}\sum_{s=1}^{64}{W^{(2)}_{t,s}}^{2}+\sum_{t=1}^{2}{b^{(2)}_{t}}^{2}\,,
\label{L2reg:CNN}
\end{eqnarray}
for the CNN. We implement our neural networks using Tensorflow~\cite{abadi2016tensorflow}.

\section{Results\label{sec:Results}}

We present our results for the ferromagnetic Ising model~\eqref{eq:Isingham}, the Blume-Capel model~\eqref{eq:IsingBC}, and the Ising-metamagnet~\eqref{eq:IsingMeta} in Sections~\ref{subsec:Isingresults},
~\ref{subsec:BCresults}, and ~\ref{subsec:MMresults} respectively. We discuss universal scaling functions and scale factors in Section~\ref{subsec:USFSF} and FSS in the vicinity of the first order transition for the Ising model~\eqref{eq:Isingham} in Section~\ref{subsec:FOI}.

\subsection{Two-dimensional Ising model}
\label{subsec:Isingresults}

 \begin{figure*}
    \centering
\includegraphics[width=\textwidth]{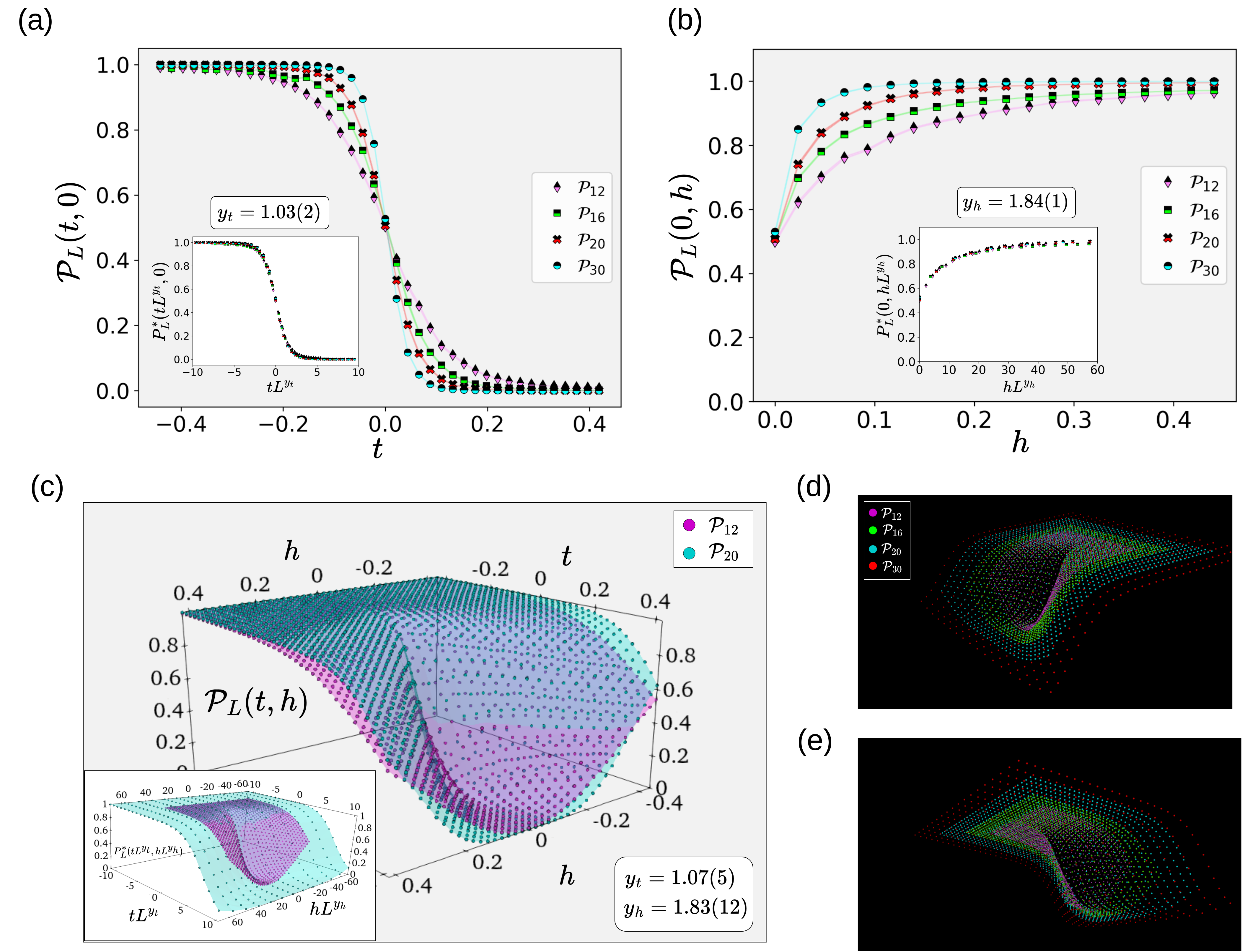}
    \caption{Plots for the 2D Ising model~\eqref{eq:Isingham}: (a) $\mathcal{P}_{L}(t,0)$ vs $t$,  for $L= 12,\,16,\,20,$ and $30$; in the inset we give the finite-size-scaling (FSS) plot of $\mathcal{P}^{*}_{L}(tL^{y_{t}} , 0)$ vs $tL^{y_t}$ from which we obtain the best-fit exponent $y_{t}=1.03(2)$ that is close to the exact value $1$. (b) $\mathcal{P}_{L}(0,h)$ vs $h$; and in the inset we give the FSS plot
of $P^{*}_{L}(0,hL^{y_{h}})$ vs $hL^{y_{h}}$; this yields the best-fit magnetic exponent $y_{h}=1.84(1)$, which is close to the exact value $1.875$. (c) $\mathcal{P}_{L}(t,h)$ as a function of $t$ and $h$ for $L=12$ and $20$; in the inset we give the FSS plot of $P^{*}_{L}(tL^{y_t},hL^{y_h})$ as a function of $tL^{y_t}$ and $hL^{y_h}$, for $L=12,16,20,30$, whence we obtain $y_{t}=1.07(5)$, and $y_{h}=1.83(12)$. In Figs.~\ref{fig:Ising_OvsTH} (d) and (e), we show different views of the surface onto which the points collapse.}
    \label{fig:Ising_OvsTH}
\end{figure*}

 \begin{figure*}
    \centering
\includegraphics[width=\textwidth]{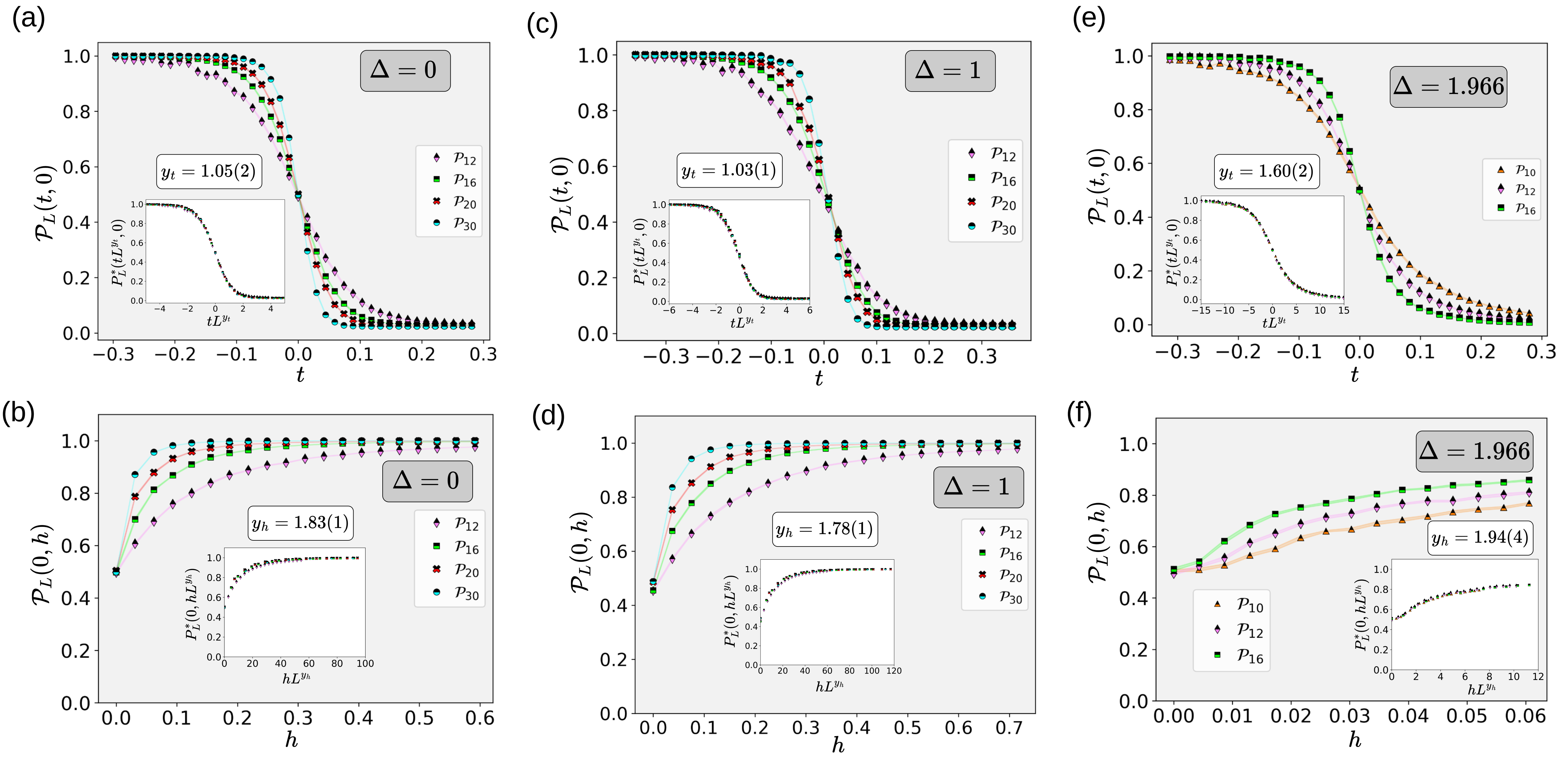}
    \caption{Plots for the 2D Blume-Capel model~\eqref{eq:IsingBC}: (a) $\mathcal{P}_{L}(t,0)$ vs $t$, for $L=12,\,16,\,20$, and $30$ for $\Delta=0$ and $h=0$; in the inset we give the finite-size-scaling (FSS) plot of $P^{*}_{L}(tL^{y_t},0)$ vs $tL^{y_t}$ from which we obtain the best-fit exponent $y_{t}=1.05(2)$ (b) $\mathcal{P}_{L}(0,h)$ vs $h$; and in the inset we give the FSS plot of $P^{*}_{L}(0,hL^{y_{h}})$ vs $hL^{y_{h}}$; this yields the best-fit magnetic exponent $y_{h}=1.83(1)$. (c) and (d) are the counterparts of (a) and (b), but with $\Delta=1$. (e) and (f) are the counterparts of (a) and (b), but in the vicinity of tricritical point at $\Delta=1.966$.}
    \label{fig:Pvs_BC}
\end{figure*}
 \begin{figure*}
    \centering \includegraphics[width=\textwidth]{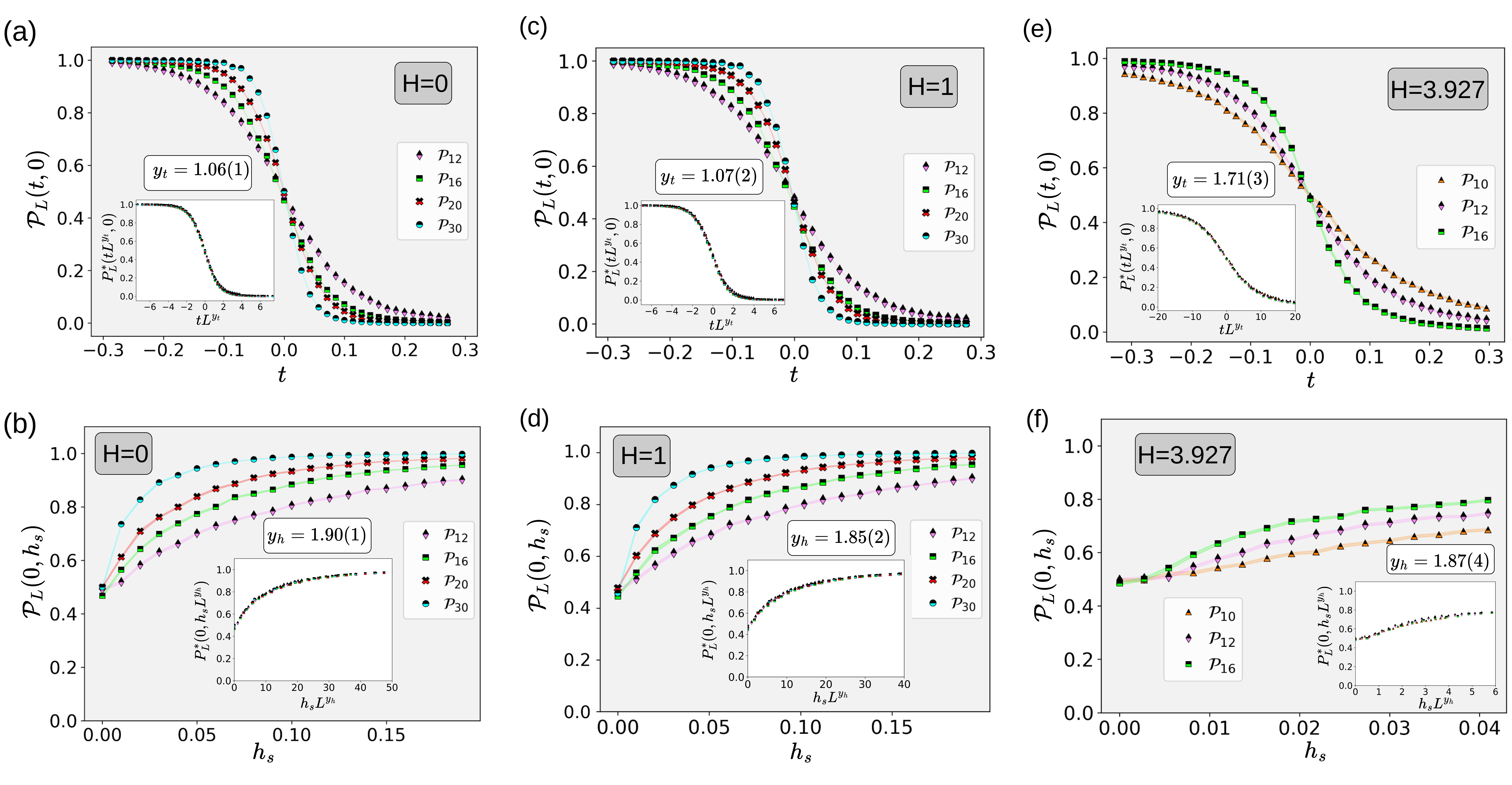}
    \caption{ Plots for the 2D Ising-metamagnet model~\eqref{eq:IsingMeta}: (a) $\mathcal{P}_{L}(t,0)$ vs $t$, for $L=12,\,16,\,20$, and $30$ for $h=0$ and $h_{s}=0$; in the inset we give the FSS plot of $P^{*}_{L}(tL^{y_t},0)$ vs $tL^{y_t}$ from which we obtain the best-fit exponent $y_{t}=1.06(1)$.
 (b) $\mathcal{P}_{L}(0,h_{s})$ vs $h_{s}$; and in the inset we give the FSS plot
of $P^{*}_{L}(0,h_{s}L^{y_{h}})$ vs $h_{s}L^{y_{h}}$; this yields the best-fit magnetic exponent $y_{h}=1.90(1)$. (c) and (d) are the counterparts of (a) and (b), but with $H=1$. (e) and (f) are the counterparts of (a) and (b), but in the vicinity of the tricritical point at $H=3.927$.}
    \label{fig:Meta}
\end{figure*}
We give the specific architectures of the CNN and FCNN that we train to classify the spin configurations of the 2D Ising model [Sec.~\ref{subsec:Ising}], above and below the critical temperature $ k_BT_c/J = 2/\ln(1+\sqrt{2})$, in  Sec.~\ref{sec:methods} [Eqs.~\eqref{eq:CNN} and \eqref{eq:FCNN}]; and the details of training and testing are given in Appendices~\ref{app:NNAT} and \ref{app:NNTest}. For a system with linear size $L$, we define 
\begin{eqnarray}
\mathcal{P}_L(t,h)&=&\langle\hat{P}_{L}(\boldsymbol{\sigma})\rangle_{\{\boldsymbol{\sigma}\}}\,,
\label{eq:PL}
\end{eqnarray}
where $\langle\cdot\rangle_{\{\boldsymbol{\sigma}\}}$ denotes the average of $\hat{P}_{L}(\boldsymbol{\sigma})$ [Eq.~\ref{eq:loss}] over the spin configurations $\boldsymbol{\sigma}$ from the test data set obtained at given values of $t$ and $h$. 
$\mathcal{P}_L(t,h)$ can be expressed as [cf. Ref.~\cite{kim2021emergence} for $h=0$]
\begin{eqnarray}
   \mathcal{P}_{L}(t,h)&=&\int\,d\boldsymbol{\sigma}\,\mathfrak{P}_{L}(\boldsymbol{\sigma},t,h)\hat{P}_L(\boldsymbol{\sigma})\,,
   \label{eq:PL1}
\end{eqnarray}
where the $\mathfrak{P}_{L}(\boldsymbol{\sigma},t,h)$ is the probabiliy distribution function (PDF) of $\boldsymbol{\sigma}$, for a system of linear size $L$; this has the FSS form~\cite{wilding1992density}
\begin{eqnarray} \mathfrak{P}_L(\boldsymbol{\sigma},t,h)&\equiv&L^{\frac{\beta}{\nu}}\mathfrak{P}_{*}(|M|L^{\frac{\beta}{\nu}},tL^{y_t},hL^{y_h})\,;
 \label{eq:PL2}
\end{eqnarray}
and $\hat{P}_{L}(\sigma)$ has the FSS form [see Ref.~\cite{kim2021emergence} and Fig.~\ref{fig:log_log} (a) in Appendix~\ref{app:Class_acc}]
\begin{eqnarray}
\hat{P}_{L}(\boldsymbol{\sigma})&\equiv&\hat{P}_{*}(|M|L^\frac{\beta}{\nu})\,;
 \label{eq:PL3}
\end{eqnarray}
using Eqs.~\eqref{eq:PL1} ~\eqref{eq:PL2} and ~\eqref{eq:PL3} we get
\begin{eqnarray}
\mathcal{P}_{L}(t,h)&=&\int d|M| L^{\frac{\beta}{\nu}}\,[\mathfrak{P}_{*}(|M|L^{\frac{\beta}{\nu}},tL^{y_t},hL^{y_h}) \nonumber \\
&\times& \hat{P}_{*}(|M|L^\frac{\beta}{\nu})]\,\nonumber\\
&\equiv&P_{*}(tL^{y_t},hL^{y_h})
\label{eq:PL4}
\end{eqnarray}  

Figure~\ref{fig:Ising_OvsTH} contains the results from our CNN~\eqref{eq:CNN}. In Fig.~\ref{fig:Ising_OvsTH} (a), we plot $\mathcal{P}_{L}(t,0)$ vs $t$, for $L=12,\,16,\,20$, and $30$; in the inset we give the finite-size-scaling (FSS) plot of $P^{*}_{L}(tL^{y_t},0)$ vs $tL^{y_t}$ from which we obtain the best-fit exponent $y_{t}=1.03(2)$ that is close to the exact value $1$ [cf. Refs.~\cite{carrasquilla2017machine, li2019extracting,zhang2019machine,kim2021emergence,shiina2020machine,chertenkov2023finite} for similar studies at $h=0$].
We now show how to generalize our discussion above to obtain the magnetic exponent $y_h$. In Fig.~\ref{fig:Ising_OvsTH} (b), we plot $\mathcal{P}_{L}(0,h)$ vs $h$; and in the inset we give the FSS plot
of $\mathcal{P}^{*}_{L}(0,hL^{y_{h}})$ vs $hL^{y_{h}}$; this yields the best-fit magnetic exponent $y_{h}=1.84(1)$, which is close to the exact value $1.875$. In Appendix~\ref{app:NNTest}, we give the estimates for $y_t$ and $y_h$, from NNs saved across different training epochs, whence we get error estimates for $y_t$ [$\simeq\,\mathcal{O}(0.01)$] and for $y_h$  [$\simeq\,\mathcal{O}(0.1)$]. Here, we use NNs, from which the estimates for $y_t$ and $y_h$ are close to the known values. The plot in Fig.~\ref{fig:Ising_OvsTH} (c) shows $\mathcal{P}_{L}(t,h)$ as a function of $t$ and $h$ for $L=12$ and $20$; in the inset we give the FSS plot of $\mathcal{P}^{*}_{L}(tL^{y_t},hL^{y_h})$ as a function of $tL^{y_t}$ and $hL^{y_h}$, for $L=12,16,20,30$, whence we obtain $y_{t}=1.07(5)$, and $y_{h}=1.83(12)$; in Figs.~\ref{fig:Ising_OvsTH} (d) and (e), we show different views of the surface onto which the points collapse. In Table.~\ref{tab_Ising}, we give the estimates for $y_t$ and $y_h$ that we obtain from scaling-collapse fits for both the CNN~\eqref{eq:CNN} and FCNN~\eqref{eq:FCNN}.   

\begin{table}[htbp] 
    \centering
    \caption{Our estimates for the thermal ($y_{t}$) and magnetic ($y_{h})$ exponents in the vicinity of the second-order transition for the
    Ising model~\eqref{eq:Isingham}}
    \label{tab_Ising}
    \begin{tabular}{|p{1.5cm}|>{\centering\arraybackslash}p{1.4cm}|>{\centering\arraybackslash}p{1.7cm}|}
        \hline
        \textit{NN}  & $y_{t}$ [exact:$1$] & $y_{h}$ [exact:$1.875$] \\
        \hline
        $\text{CNN}$  & $1.03(2)$ &  $1.84(1)$ \\
        \hline
        $\text{FCNN}$  & $1.06(2)$ &  $1.85(1)$ \\
        \hline
    \end{tabular}
\end{table}

\subsection{Two-dimensional Blume-Capel model} 
\label{subsec:BCresults}

We train our CNN~\eqref{eq:CNN} and FCNN~\eqref{eq:FCNN} to classify the spin configurations of the 2D Blume-Capel model~\eqref{eq:IsingBC} above and below its critical temperature $T_c\simeq 1.69$~\cite{mozolenko2024blume}, with $\Delta=0$ and $h=0$ [details of the training and testing are given in Appendices~\ref{app:NNAT} and ~\ref{app:NNTest}]. 

Figure~\ref{fig:Pvs_BC} contains the results from our CNN~\eqref{eq:CNN}.
In Fig.~\ref{fig:Pvs_BC} (a), we plot $\mathcal{P}_{L}(t,0)$ vs $t$, for $L=12,\,16,\,20$, and $30$ for $\Delta=0$ and $h=0$; in the inset we give the finite-size-scaling (FSS) plot of $\mathcal{P}^{*}_{L}(tL^{y_t},0)$ vs $tL^{y_t}$ from which we obtain the best-fit exponent $y_{t}=1.05(2)$.
In Fig.~\ref{fig:Pvs_BC} (b), we plot $\mathcal{P}_{L}(0,h)$ vs $h$; and in the inset we give the FSS plot of $\mathcal{P}^{*}_{L}(0,hL^{y_{h}})$ vs $hL^{y_{h}}$; this yields the best-fit magnetic exponent $y_{h}=1.83(1)$. In Table~\ref{tab:BC_so}, we give the estimates for $y_t$ and $y_h$ that we obtain from scaling-collapse fits for both the CNN~\eqref{eq:CNN} and FCNN~\eqref{eq:FCNN}.

We now test our CNN~\eqref{eq:CNN}, trained at $\Delta=0$ and $h=0$, for FSS along the second-order transition line; for the 2D Blume-Capel model~\eqref{eq:IsingBC}; this transition is in the universality class of the 2D Ising model~\cite{beale1986finite,kwak2015first,moueddene2024critical}. In particular, in Fig.~\ref{fig:Pvs_BC} (c), we fix $\Delta=1$, and change $T$,  in the vicinity of $\mathcal{P}_{L}(t,h=0)\simeq0.5$. Now FSS best-fit plots [insets in Figs.~\ref{fig:Pvs_BC} (c) and (d)] yield $y_{t}=1.03(1)$ and $y_{h}=1.78(1)$. We summarize these results for the CNN~\eqref{eq:CNN} and FCNN~\eqref{eq:FCNN} in Table~\ref{tab:BC_so}. 

\begin{table}[htbp] 
    \centering
    \caption{Our estimates for the thermal ($y_{t}$) and magnetic ($y_{h})$ exponents in the vicinity of the second-order transition for $\Delta=0$ and $\Delta=1$ in the 
    Blume-Capel model~\eqref{eq:IsingBC}.}
    \label{tab:BC_so}
    \begin{tabular}{|p{2.3cm}|>{\centering\arraybackslash}p{1.4cm}|>{\centering\arraybackslash}p{1.7cm}|}
        \hline
        \textit{NN}  & $y_{t}$ [exact:$1$] & $y_{h}$ [exact:$1.875$] \\
        \hline
        \text{CNN}: $\Delta=0$& $1.05(2)$&  $1.83(1)$\\
        \hline
        \text{FCNN}: $\Delta=0$& $1.07(2)$ &  $1.81(1)$\\
        \hline 
        \text{CNN}: $\Delta=1$& $1.03(1)$&  $1.78(1)$\\
        \hline
        \text{FCNN}: $\Delta=1$& $1.09(1)$ & $2.0(1)$ \\
        \hline 
    \end{tabular}
\end{table}

The Blume-Capel model~\eqref{eq:IsingBC} exhibits a tricritical point at  $\Delta\simeq1.966$ and $T\simeq0.608$ [see, e.g., Ref.~\cite{kwak2015first}]; the exponents at this tricritical point are different from those that characterize the 2D Ising universality class~\cite{beale1986finite,wilding1996tricritical,kwak2015first,moueddene2024critical}. Here, we train our CNN~\eqref{eq:CNN} and FCNN~\eqref{eq:FCNN} to classify the spin configurations below and above $T_c=0.608$~\cite{kwak2015first}, with $\Delta=1.966$. In Fig.~\ref{fig:Pvs_BC} (e), we plot $\mathcal{P}_{L}(t,0)$ vs $t$, for $L=10,\,12,\,16$; in the inset we give the FSS plot of $\mathcal{P}^{*}_{L}(tL^{y_t},0)$ vs $tL^{y_t}$, whence we obtain the best-fit exponent $y_{t}=1.60(2)$, which is close to $y_{t}\simeq1.8$~\cite{beale1986finite,wilding1996tricritical,kwak2015first,moueddene2024critical}.
In Fig.~\ref{fig:Pvs_BC} (f), we plot $\mathcal{P}_{L}(0,h)$ vs $h$; and in the inset we give the FSS plot
of $\mathcal{P}^{*}_{L}(0,hL^{y_{h}})$ vs $hL^{y_{h}}$; this yields the best-fit magnetic exponent $y_{h}=1.94(4)$, which is close to $y_{h}\simeq1.92$~\cite{beale1986finite,wilding1996tricritical,kwak2015first,moueddene2024critical}. In Table.~\ref{tab:BC_tri}, we give the estimates for $y_t$ and $y_h$ that we obtain from scaling-collapse fits for both the CNN~\eqref{eq:CNN} and FCNN~\eqref{eq:FCNN}.

\begin{table}[htbp] 
    \centering
    \caption{Our estimates for the thermal ($y_{t}$) and magnetic ($y_{h})$ exponents in the vicinity of the tri-critical point for the Blume-Capel model~\eqref{eq:IsingBC}.}
    \label{tab:BC_tri}
    \begin{tabular}{|p{2.8cm}|>{\centering\arraybackslash}p{1.4cm}|>{\centering\arraybackslash}p{1.7cm}|}
        \hline
        \textit{NN}  & $y_{t}$ [exact:$1.8$] & $y_{h}$ [exact:$1.92$] \\
        \hline
        \text{CNN}: $\Delta=1.966$ & $1.60(2)$&  $1.94(4)$\\
        \hline
        \text{FCNN}: $\Delta=1.966$& $1.87(3)$ &  $1.94(3)$\\
        \hline 
    \end{tabular}
\end{table}

\subsection{Two-dimensional Ising-metamagnet model}
\label{subsec:MMresults}
We turn now to the 2D Ising-metamagnet model~\eqref{eq:IsingMeta}. We train our CNN~\eqref{eq:CNN} and FCNN~\eqref{eq:FCNN} to classify the staggered spin configurations of the 2D Ising-metamagnet model~\eqref{eq:IsingMeta} above and below its critical temperature $T_c\simeq5.263$~\cite{beale1984finite}, with $h=0$ and $h_{s}=0$ [details of the training and testing are given in Appendices~\ref{app:NNAT} and ~\ref{app:NNTest}]. 

Figure~\ref{fig:Meta} contains the results from our CNN~\eqref{eq:CNN}.
In Fig.~\ref{fig:Meta} (a), we plot $\mathcal{P}_{L}(t,0)$ vs $t$, for $L=12,\,16,\,20$, and $30$ for $h=0$ and $h_{s}=0$; in the inset we give the FSS plot of $P^{*}_{L}(tL^{y_t},0)$ vs $tL^{y_t}$ from which we obtain the best-fit exponent $y_{t}=1.06(1)$.
In Fig.~\ref{fig:Meta} (b), we plot $\mathcal{P}_{L}(0,h_{s})$ vs $h_{s}$; and in the inset we give the FSS plot
of $\mathcal{P}^{*}_{L}(0,h_{s}L^{y_{h}})$ vs $h_{s}L^{y_{h}}$; this yields the best-fit magnetic exponent $y_{h}=1.90(1)$. In Table~\ref{tab:Meta}, we give the estimates for $y_t$ and $y_h$ that we obtain from scaling-collapse fits for both the CNN~\eqref{eq:CNN} and FCNN~\eqref{eq:FCNN}.

We now test our CNN~\eqref{eq:CNN}, trained at $h=0$ and $h_{s}=0$, for FSS along the second-order transition line; for the 2D Ising-metamagnet model~\eqref{eq:IsingMeta}, this transition is in the universality class of the 2D Ising model~\cite{beale1984finite}. In particular, in Fig.~\ref{fig:Meta} (c), we fix $H=1$, and change $T$,  in the vicinity of $\mathcal{P}_{L}(t,h=0)\simeq0.5$. Now the FSS best-fit plots [insets in Figs.~\ref{fig:Meta} (c) and (d)] yield $y_{t}=1.07(2)$ and $y_{h_{s}}=1.85(2)$. We summarize these results for the CNN~\eqref{eq:CNN} and FCNN~\eqref{eq:FCNN} in Table~\ref{tab:Meta}.

The 2D Ising-metamagnet model~\eqref{eq:IsingMeta} exhibits a tricritical point at $h_{s}=0$, $H\simeq3.927$, and $T\simeq2.41$ [see, e.g., ~\cite{beale1984finite}]; the exponents at this tricritical point~\cite{beale1984finite,landau1981tricritical,herrmann1984finite} are different from those that characterize the 2D Ising universality class. Here, we train our CNN~\eqref{eq:CNN} and FCNN~\eqref{eq:FCNN} to classify the spin configurations below and above $T_c=2.41$~\cite{beale1984finite}, with $H_{s}=0$ and $H=3.927$. In Fig.~\ref{fig:Meta} (e), we plot $\mathcal{P}_{L}(t,0)$ vs $t$, for $L=10,\,12,\,16$; in the inset we give the FSS plot of $\mathcal{P}^{*}_{L}(tL^{y_t},0)$ vs $tL^{y_t}$, whence we obtain the best-fit exponent $y_{t}=1.71(3)$, which is close to $y_{t}\simeq1.8$\cite{beale1984finite}.
In Fig.~\ref{fig:Meta} (f), we plot $\mathcal{P}_{L}(0,h_{s})$ vs $h_{s}$; and in the inset we give the FSS plot
of $P^{*}_{L}(0,h_{s}L^{y_{h}})$ vs $h_{s}L^{y_{h}}$; this yields the best-fit magnetic exponent $y_{h}=1.87(4)$, which is close to $y_{h}\simeq1.92$. In Table~\ref{tab:Meta_Tri}, we give the estimates for $y_t$ and $y_h$ that we obtain from scaling-collapse fits for both the CNN~\eqref{eq:CNN} and FCNN~\eqref{eq:FCNN}.

\begin{table}[htbp] 
    \centering
    \caption{Our estimates for the thermal ($y_{t}$) and magnetic ($y_{h})$ exponents in the vicinity of the second-order transitions for $H=0$ and $H=1$ for the Ising-metamagnet model~\eqref{eq:IsingMeta}.}
    \label{tab:Meta} 
\begin{tabular}{|p{2.3cm}|>{\centering\arraybackslash}p{1.5cm}| >{\centering\arraybackslash}p{1.8cm}|}
        \hline
        \textit{NN}  & $y_{t}$ 
  [exact:$1$] & $y_{h}$ [exact:$1.875$]\\
        \hline
        \text{CNN}: $H=0$& $1.06(1)$&  $1.90(1)$\\
        \hline
        \text{FCNN}: $H=0$& $1.09(1)$ &  $1.85(2)$\\
        \hline 
        \text{CNN}: $H=1$& $1.07(2)$&  $1.85(2)$\\
        \hline
        \text{FCNN}: $H=1$& $1.092(1)$ &  $1.93(2)$\\
        \hline 
    \end{tabular}
\end{table}


\begin{table}[htbp] 
    \centering
    \caption{Our estimates for the thermal ($y_{t}$) and magnetic ($y_{h})$ exponents in the vicinity of the tricritical point  for the Ising-metamagnet model~\eqref{eq:IsingMeta}.}
    \label{tab:Meta_Tri}
    \begin{tabular}{|p{2.3cm}|>{\centering\arraybackslash}p{1.8cm}| >{\centering\arraybackslash}p{1.8cm}|}
        \hline
        \textit{NN}  & $y_{t}$ [exact:$1.80$] & $y_{h}$ [exact:$1.92$] \\
        \hline
        \text{CNN}& $1.71(3)$&  $1.87(4)$\\
        \hline
        \text{FCNN}& $1.69(2)$ &  $2.03(5)$\\
        \hline 
    \end{tabular}
\end{table}

\begin{figure*}
    \centering \includegraphics[width=\textwidth]{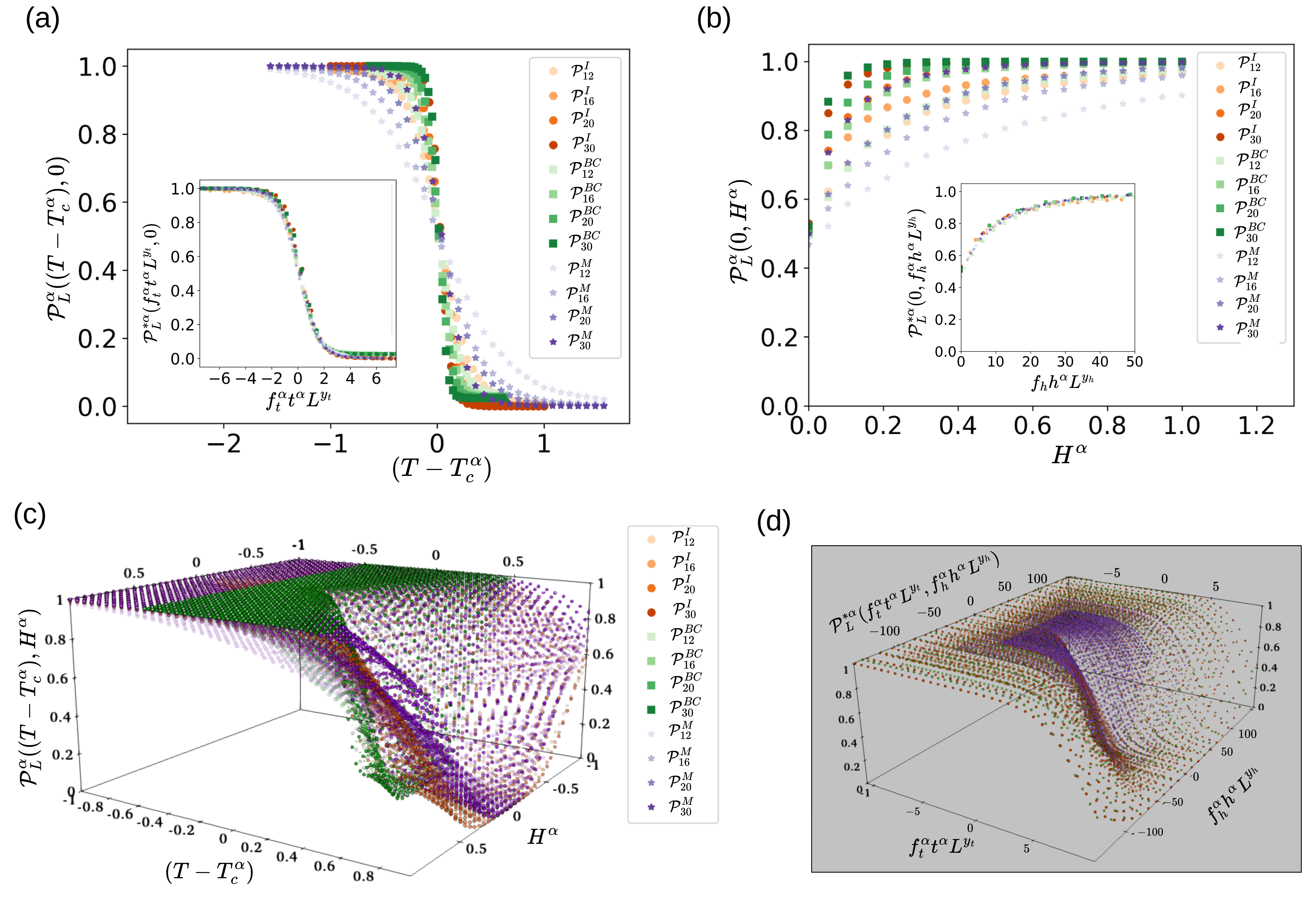}
    \caption{(a) Plots of $P^I_{L}((T-T_{c}^{I}),0)$ vs $(T-T_{c}^{I})$ , $P^{BC}_{L}((T-T_{c}^{BC}),0)$ vs $(T-T_{c}^{BC})$, and $P^M_{L}((T-T_{c}^{M}),0)$ vs $(T-T_{c}^{M})$, for $L=12,\,16,\,20$, and $30$; these curves are different for the three models we consider. (b) $P^I_{L}(0,H)$ vs $H$, $P^{BC}_{L}(0,H)$  vs $H$, and $P^M_{L}(0,H_{s})$ vs $H_{s}$, for $L=12,\,16,\,20$,
and $30$. (c) $P^\alpha_{L}((T-T_{c}^{\alpha}),H)$ and $P^\alpha_{L}((T-T_{c}^{\alpha}),H_{s})$ vs $T-T_{c}^{\alpha}$ and $H$ or $H_{s}$, for all three models, $\alpha = I,\,BC,\,M$ . (d) These surfaces collapse onto one universal scaling surface, if we use the scaled variables $f^{BC}_ttL^{y_{t}}$ and $f^{BC}_hhL^{y_{h}}$, for the Blume-Capel model~\eqref{eq:IsingBC}, and $f^{M}_ttL^{y_{t}}$ and $f^{M}_hh_{s}L^{y_{h}}$, for the Ising-metamagnet model~\eqref{eq:IsingMeta}.}
    \label{fig:two_scale}
\end{figure*}

 \begin{figure*}
    \centering \includegraphics[width=\textwidth]{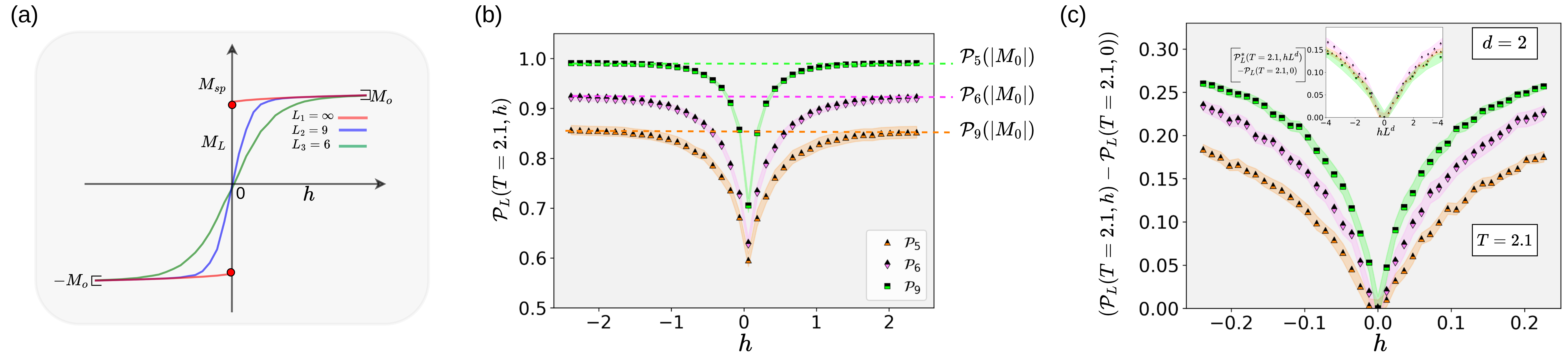}
    \caption{(a) Plots of $M_L$ vs $h$ for the Ising first-order transition in the vicinity of $T_c=2.1$.  (b) $\mathcal{P}_{L}(T=2.1,h)$ vs $h$ (c) $(\mathcal{P}_{L}(T=2.1,h)-\mathcal{P}_{L}(T=2.1,h=0))$ vs $h$, and in the inset is the collapse plot of $(\mathcal{P}^{*}_{L}(T=2.1,hL^{d})-\mathcal{P}_{L}(T=2.1,h=0))$ vs $hL^{d}$.}
    \label{fig:Ising_FO}
\end{figure*}

\subsection{Universal scaling functions and scale factors\label{subsec:USFSF}} We now investigate two-scale factor universality~\cite{PhysRevB.9.2107,PhysRevB.13.2986,PhysRevLett.29.345} by comparing the scaling functions $\mathcal{P}_{L}^{*\alpha}(tL^{y_t},hL^{y_h})$ obtained from our neural networks, where the label $\alpha$ is $I,\, BC$, and $M$ for the 2D Ising~\eqref{eq:Isingham}, 2D Blume-Capel~\eqref{eq:IsingBC}, and 2D Ising-metamagnet~\eqref{eq:IsingMeta} models, respectively. The plots in Fig.~\ref{fig:two_scale} (a) of $\mathcal{P}^I_{L}((T-T_{c}^{I}),0)$ vs $(T-T_{c}^{I})$, $\mathcal{P}^{BC}_{L}((T-T_{c}^{BC}),0)$ vs $(T-T_{c}^{BC})$, and $\mathcal{P}^M_{L}((T-T_{c}^{M}),0)$ vs $(T-T_{c}^{M})$, for $L=12,\,16,\,20$,
and $30$ show that these curves are different for these three models. We now show that, if we scale  $tL^{y_{t}}$ by model-dependent scale factors, then all these curves collapse onto one universal curve. In particular, if we use the scaled variables  
\begin{equation}
f^{BC}_ttL^{y_{t}}\;\; {\rm{and}} \;\; f^{M}_ttL^{y_{t}} 
\label{eq:fBCfM}
\end{equation}
for the Blume-Capel and the Ising-metamagnet models, respectively, then we get the universal curve shown in the inset of Fig.~\ref{fig:two_scale} (a) and the best fit yields the model-dependent scale factors $f_t^{BC} \simeq 1.27(6)$ and $f_t^{M} \simeq 1.1(4)$.  Similarly, the plots in Fig.~\ref{fig:two_scale} (b) of $P^I_{L}(0,H)$ vs $H$, $P^{BC}_{L}(0,H)$  vs $H$, and $P^M_{L}(0,H_{s})$ vs $H_{s}$, for $L=12,\,16,\,20$,
and $30$, demonstrate that these curves are different for these three models; but, if we use the scaled variables  
\begin{equation}
f^{BC}_hhL^{y_{h}}\;\; {\rm{and}} \;\; f^{M}_hh_{s}L^{y_{h}} 
\label{eq:fBCfMh}
\end{equation}
for the Blume-Capel and the Ising-metamagnet models, respectively, then we get the universal curve shown in the inset of Fig.~\ref{fig:two_scale} (b) and the best fit yields the model-dependent scale factors $f_h^{BC} \simeq 0.95(10)$ and $f_h^{M} \simeq 1.02(5)$. The plots in the insets of Figs.~\ref{fig:two_scale} (a) and (b) are a result of two-scale factor universality~\cite{PhysRevB.9.2107,PhysRevB.13.2986,PhysRevLett.29.345} in our NN study. In Fig.~\ref{fig:two_scale} (c), we give plots of $P^\alpha_{L}((T-T_{c}^{\alpha}),H\,\text{or}H_{s})$ vs $T-T_{c}^{\alpha}$ and $H$ or $H_{s}$, for all three models, i.e., $\alpha$ is $I,\, BC$, and $M$; these surfaces collapse onto one universal scaling surface [Fig.~\ref{fig:two_scale} (d)], if we use the scaled variables $f^{BC}_ttL^{y_{t}}$ and $f^{BC}_hhL^{y_{h}}$, for the Blume-Capel model, and $f^{M}_ttL^{y_{t}}$ and $f^{M}_hh_sL^{y_{h}}$, for the Ising-metamagnet model. 

If $N^{\alpha}_{L}$ is the number of spin configurations that are correctly classified out of a total $N$ configurations for a system of linear size $L$, say below $T^{\alpha}_{c}$ and in the temperature range $[T_{0},T^{\alpha}_{c}]$, then we can show [see Appendix~\ref{app:Class_acc}]
\begin{eqnarray}
\frac{(N-N^{\alpha}_{L})}{N}\propto\frac{1}{f_{t}^{\alpha}|t^{\alpha}_{0}|L^{y_t}}\,,
\label{eq:class}
\end{eqnarray}  
where $t^{\alpha}_{0}\equiv\frac{T_{0}-T^{\alpha}_{c}}{T_{c}}$. The plot of $\frac{(N-N^{\alpha}_{L})}{N}$ vs $f^{\alpha}|t^{\alpha}_{0}|L^{y_t}$ in Fig.~\ref{fig:log_log} (b) in Appendix~\ref{app:Class_acc}  verifies this relation.

\subsection{First-order phase boundary in the 2D Ising model\label{subsec:FOI}}

To investigate the FSS of the CNNs in the vicinity of  the first-order transition, we follow the theoretical ideas given in Refs.~\cite{nienhuis1976renormalization,klein1976essential,fisher1982scaling,binder1984finite,berker1976blume}. In a renormalization-group (RG) treatment, the jump of the order parameter at a first-order transition is governed by the flows of the RG recursion relations in the vicinity of a discontinuity fixed point [see, e.g., Ref.~\cite{berker1976blume}]; such a fixed point occurs in the sub-space of even-spin couplings (i.e., $h$ and all odd-spin couplings vanish) and lies out at infinity (i.e., at $J/(k_BT)$ and many other even-spin couplings $=\infty$)~\footnote{A non-perturbative RG, of the type used in Ref.~\cite{berker1976blume}, is essential for obtaining such fixed points that lie at infinite values of couplings.}; the dominant eigenvalues at this discontinuity fixed point are such that $y_h =d$ and $y_t=(d-1)$, where $d$ is the spatial dimension~\cite{nienhuis1976renormalization,klein1976essential,berker1976blume}. Therefore, in the vicinity of the first-order transition~\cite{binder1984finite} in the Ising model~\eqref{eq:Isingham} we have the following: for $\frac{HM_{sp}L^{d}}{k_{b}T}\gg1$, where $M_{sp}$ is the infinite-system magnetization at $h\to0^{+}$~\cite{binder1984finite}], the magnetization for the system of linear size $L$ is $M_{L}$, which $\to M_{0}$, the equilibrium magnetization value for the infinite $L$ and $h$ [see Fig.~\ref{fig:Ising_FO} (a)]. If $\frac{HM_{sp}L^{d}}{k_{b}T}\ll1$, we have $M_{L} \propto h L^{d}$ ~\cite{binder1984finite}.

Transfer learning allows us to use the CNNs, which we have trained in the vicinity of the critical point of the Ising model~\eqref{eq:Isingham}, to obtain plots of $P_{L}(T=2.1,h)$,  as we change $h$ from positive to negative values to cross the first-order boundary at $T=2.1 < T_c$ [Fig.~\ref{fig:Ising_FO} (b)]; we show plots for $L=5,\,6$ and $9$. From these plots we see that $\mathcal{P}_{L}$ tends to $\mathcal{P}_{L}(|M_{0}|)$, for large values of $h$; and $\mathcal{P}_{L}(|M|_L(h=0))$ at $h=0$ depends on the value of $L$. In Fig.~\ref{fig:Ising_FO} (c), we present plots of $(\mathcal{P}_L(T=2.1,h)-\mathcal{P}_L(T=2.1,h=0))$ vs $h$; in the inset we show that these curves collapse onto one curve if we plot $(\mathcal{P}^{*}_L(T=2.1,hL^{d})-\mathcal{P}_L(T=2.1,h=0))$ vs $hL^{d}$, with $d=2$.

\section{\label{sec:conclusions}Discussion and Conclusions}

 Earlier studies~\cite{carrasquilla2017machine,  li2019extracting,kim2021emergence,shen2021supervised,shiina2020machine,chertenkov2023finite,wang2024supervised} of neural-network-aided methods for the determination of critical exponents have considered only simple critical points, such as the one in the Ising model~\eqref{eq:Isingham} at $h=0$. Our work goes well beyond these earlier investigations by developing a full framework for the scaling forms of NN outputs in the vicinities of critical and tricritical points and obtaining the following: (a) both the thermal and magnetic exponents, $y_t$ and $y_h$, not only at simple Ising-type critical points, but also at tricritical points in the Blume-Capel~\eqref{eq:IsingBC} and the Ising-metamagnet~\eqref{eq:IsingMeta} models; (b) the full scaling form for the NN outputs $\mathcal{P}_L$ given in Eq.~\eqref{eq:PL4}; (c) the non-universal scale factors that are required for two-scale-factor universality; and (d) the CNN manifestation of finite-size scaling at the Ising-model first-order phase boundary.

From the machine-learning point of view, our study provides the following interesting applications of transfer learning: We first train our CNNs and FCNNs in the vicinities of the zero-field critical points, at $h=0$ of the Ising model~\eqref{eq:Isingham}, at $\Delta=0,\,h=0$ in the Blume-Capel model~\eqref{eq:IsingBC}, and at $h_s=0,\,h=0$ for the Ising-metamagnet~\eqref{eq:IsingMeta}; and then we use these trained NNs to uncover scaling and critical exponents at critical points that occur at $h\neq0$ for the Ising model~\eqref{eq:Isingham}, $h\neq0$ and $\Delta\,\neq0$ for the Blume-Capel~\eqref{eq:IsingBC} model, and $h\neq0,\,h_s\neq0$ for the Ising-metamagnet model~\eqref{eq:IsingMeta}. Our studies are valuable because they help us to go beyond the mere classification of phase by understanding the behavior and characteristics of the NN outputs $\mathcal{P}_L(t,h)$ and then extract, from these characteristics, other useful quantities, 
e.g., the scaling form for the number of spin configurations that are correctly classified [see Eq.~\eqref{eq:class}].

The methods that we have developed can be extended, via transfer learning, to study phase transitions in other models. For example, the well-known Ising-lattice-gas mapping relates the Ising spins $S_i$ to lattice-gas variables $n_i = [S_i+1]/2$; given that $S_i=\pm1$, we have $n_i=0\; {\rm{or}} \; 1$ and the up-spin (down-spin) of the Ising model maps onto the high-density (low-density) phase of the lattice gas; in general, the liquid-gas critical point in a $d$-dimensional continuum fluid should be in the universality class of the $d$-dimensional Ising model. Therefore, it is interesting to use our trained NNs first to study Ising-model criticality in $d=3$ and then employ transfer learning to examine the liquid-gas critical point in a continuum fluid described, e.g., the Lennard-Jones potential~\cite{frenkel2023understanding,goldenfeld2018lectures}
as we will show in future work.

Although machine-learning methods, e.g., those that we have employed here, have, so far, not yielded values of critical exponents that are as accurate as those obtained by 
high-resolution MC, conformal-bootstrap, and series-expansions methods [e.g., see Refs.~\cite{ferrenberg2018pushing,xu2020high,butera2002critical,blote1999cluster,deng2003simultaneous,ozeki2007nonequilibrium,weigel2010error,hasenbusch2010finite,kos2016precision,wang2014phase} for the three-dimensional Ising model]. We expect that, as these ML methods are refined, by building on the framework that we have outlined here, the ML-based determination of critical exponents will move apace and achieve accuracies comparable to
those attained by the other methods mentioned above. 

\begin{acknowledgments}  
\hspace{10pt} We thank M. Barma for discussions, the Anusandhan National Research Foundation (ANRF), the Science and Engineering Research Board (SERB), and the National Supercomputing Mission (NSM), India, for support,  and the Supercomputer Education and Research Centre (IISc), for computational resources. 
\end{acknowledgments}

\section*{Data and code availibility}
\hspace{10pt}The data and code utilized in this study can be made available from the authors upon reasonable request.

\section{Appendices}
\label{app}

In Appendix~\ref{app:NNAT} we give the details of the training data-sets and neural-network training. This is followed by Appendix~\ref{app:NNTest}, in which we give the details of the data-sets and neural networks that we use for testing and error estimates. Finally, in Appendix~\ref{app:Class_acc} we give a detailed derivation of Eq.~\eqref{eq:class}.

\subsection{\label{app:NNAT} Data-sets and neural network training}

For the 2D Ising model~\eqref{eq:Isingham} in the vicinity of second-order transition,  we carry out Monte Carlo (MC) simulations, for systems with linear sizes $L=12,\,16,\,20,\,{\rm{and}}\,30$, at $100$ different values of the temperature $T$, spaced at intervals of $0.01$ in the range $[T_{c}-0.5,T_{c}+0.5]$, where $T_c \simeq 2.269$ is the critical temperature. For each value of $T$, we discard the first  $10^{6}$ Monte Carlo steps  per spin (MCS/S) and include $2000$ spin configurations, from the subsequent $10^{6}$ MCS/S; thus, we obtain a total of $2\times10^{5}$
spin-configuration snapshots, which we then use to train our neural networks (NNs). We carry out similar MC simulations to obtain training data for the Blume-Capel model~\eqref{eq:IsingBC}, with training data in the range $[T_{c}-0.5,T_{c}+0.5]$, where $T_{c}\simeq1.693$,  and for the Ising-metamagnet model~\eqref{eq:IsingMeta}, with training data in the range $[T_{c}-1,T_{c}+1]$, where $T_{c}\simeq5.263$. For the 2D Blume-Capel model~\eqref{eq:IsingBC} and the Ising-metamagnet model~\eqref{eq:IsingMeta}, while obtaining the spin configurations to train NNs in the vicinity of their tricritical points, we discard the first $10^{7}$ MCS/S, and include $2000$ spin configurations from the next $10^{7}$ steps for each of the $100$ values of temperature in the ranges $[T_c-0.125,T_c+0.125]$ ($T_c\simeq0.608$) and $[T_c-0.75,T_c+0.75]$ ($T_c\simeq2.41$), respectively. 

To test the NNs in the vicinity of the first-order transition for the 2D Ising model~\eqref{eq:Isingham}, we train systems of linear size $L=\,5,\,6,\,9$, where we obtain the training spin configurations from MC simulations, in the temperature range $[T_c-1,T_c+1]$ [see discussion above].

We train our NNs for $9000-10000$ epochs and save the NNs, after every 100 epochs; finally, we obtain 10 NNs, for each system size, which we use on our test data set [ Section~\ref{app:NNTest}]. We train our NNs with the Adam Optimizer~\cite{kingma2014adam}, with an initial learning rate of $10^{-3}$ and a batch size of $256$.
To prevent the overfitting of NNs, we set the regularizer strength  $\lambda=0.005$ in Eq.~\eqref{eq:loss}.

\begin{figure*}
    \centering  \includegraphics[width=\textwidth]{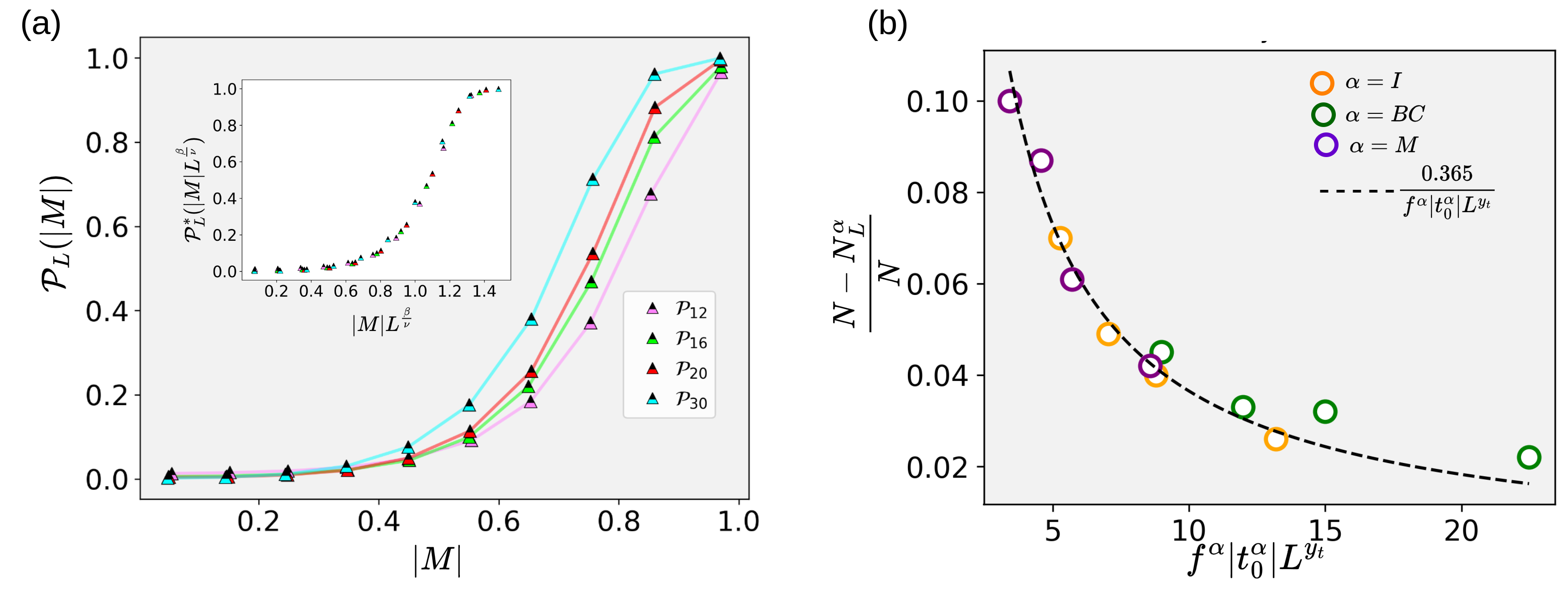}
    \caption{(a) Plot of $\mathcal{P}_{L}(|M|)$ vs $|M|$, and in the inset, a plot of $\mathcal{P}_{L}^{*}(|M|L^{\frac{\beta}{\nu}})$ vs $|M|L^{\frac{\beta}{\nu}}$, for the Ising model~\eqref{eq:Isingham}. (b) Plot of 
$\frac{N-N_{L}^{\alpha}}{N}$ vs $f^{\alpha}|t^{\alpha}_{0}|L^{y_t}$ [see Eq.~\eqref{eq:class_App}] for the Ising ($I$)~\eqref{eq:Isingham}, Blume-Capel ($BC$)~\eqref{eq:IsingBC}, and Ising-metamagnet ($M$)~\eqref{eq:IsingMeta} models.}
    \label{fig:log_log}
\end{figure*}

\subsection{\label{app:NNTest}Neural networks and data-sets for testing and error estimations}

In the vicinity of a second-order transition, we calculate $P_L$ for the test data  as follows: We discard the first $10^{6}$ MCS/S, at each temperature, and consider $5000$ snapshots from the next $10^{6}$ MCS/S. We draw $2000$ snapshots randomly, $5$ times, from these $5000$ snapshots. This yields $5$ values of $P_L$, per NN, so, for $10$ NNs [see Sec.~\ref{app:NNAT}], we obtain $50$ values
from which we obtain the mean and the error for $P_{L}$.  From the $50$ estimates for $P_{L}$, obtained for the Ising model~\eqref{eq:Isingham}, we obtain $y_t=1.06(7)$, and $y_h=1.73(26)$ for the CNNs. Similarly, for the FCNNs, we obtain $y_t=1.0(3)$ and $y_h=1.80(30)$. In Fig.~\ref{fig:Ising_OvsTH} we use the CNN set for which $\mathcal{P}_{L}(0,0)$ is close to $0.5$. In Table~\ref{tab_Ising}, we use the NN set that gives the estimates of $y_t$ and $y_h$ which are close to the known values; to obtain the error estimates for this NN set, we draw $2000$ spin-configurations randomly, from the $5000$ spin configurations; we repeat this $10$ times, and obtain the mean and the standard deviations for the best fits for the exponents [we use the same procedure for error estimation for all the models in the vicinity of a second-order transition]. 

We use the procedure, described in the previous paragraph, for the Blume-Capel~\eqref{eq:IsingBC} model at $\Delta=0$. This yields $y_{t}=1.05(6)$ and $y_h=1.91(30)$, for the CNNs, and $y_{t}=1.06(4)$ and $y_h=1.77(25)$, for the FCNNs. In Figs.~\ref{fig:Pvs_BC} (a) and (b), we used the CNNs for which $P_L(0,0)$ is close to $0.5$. In Table~\ref{tab:BC_so}, we use the CNN and FCNN sets that yield estimates close to known values of $y_t$ and $y_h$. For $\Delta=1$, when we use the CNNs trained for $\Delta=0$, this yields $y_t=1.05(5)$ and $y_h=1.76(23)$; similarly, for FCNNs, we obtain $y_t=1.03(9)$ and $y_h=2.00(30)$. For $\Delta=1$, in Figs.~\ref{fig:Pvs_BC} (c) and (d), and Table~\ref{tab:BC_so}, we use the CNN and FCNN set that we have used for $\Delta=0$ [Figs.~\ref{fig:Pvs_BC} (a) and (b)].

A similar procedure for the  Ising metagmagnet~\eqref{eq:IsingMeta} model at $H=0$ yields $y_{t}=1.07(4)$ and $y_h=1.74(27)$, for the CNNs, and $y_{t}=1.08(4)$ and $y_h=1.85(39)$, for the FCNNs. In Figs.~\ref{fig:Meta} (a) and (b), and Table~\ref{tab:Meta}, we use the CNN and the FCNN sets that yield estimates close to known values of $y_t$ and $y_h$. For $H=1$, the CNNs trained with $H=0$ yield $y_t=1.09(3)$ and $y_h=1.70(30)$; similarly, for the FCNNs, we obtain $y_t$ = $1.1(5)$ and $y_h=1.89(31)$. In Figs.~\ref{fig:Meta} (c) and (d), and Table.~\ref{tab:Meta}, for $H=1$, we use the NN set that we employ for $H=0$ [Figs.~\ref{fig:Meta} (a) and (b)].

In the vicinity of the tricritical point, we calculate $P_L$ for the test data as follows: We discard the first $10^{7}$ MCS/S, at each temperature, and consider $5000$ snapshots from the next $10^{7}$ MCS/S. We draw $2000$ snapshots randomly, $5$ times, from these $5000$ snapshots. This yields $5$ values of $P_L$, per NN, so, for $10$ NNs [see Sec.~\ref{app:NNAT}], we obtain $50$ values
from which we obtain the mean and the error for $P_{L}$.  From the $50$ estimates for $P_{L}$, obtained for the Blume-Capel model~\eqref{eq:IsingBC}, we obtain $y_t=1.51(8)$, and $y_h=1.96(44)$ for the CNNs. Similarly, for the FCNNs, we obtain $y_t=2.00(15)$ and $y_h=2.08(27)$. In Figs.~\ref{fig:Pvs_BC} (e) and (f), and Table~\ref{tab:BC_tri}, we use the NN set that gives the estimates of $y_t$ and $y_h$ which are close to the known values; to obtain the error estimates for this NN set, we draw $2000$ spin-configurations randomly, from the $5000$ spin configurations; we repeat this $50$ times, and obtain the mean and the standard deviations for the best fits for the exponents [we use the same procedure for error estimation for the Ising-metamagnet model~\eqref{eq:IsingMeta} in the vicinity of its tricritical point]. 

We use the procedure, described in the previous paragraph for the Ising-metamagnet~\eqref{eq:IsingMeta} model in the vicinity of the tricritical point. This yields $y_{t}=1.66(3)$ and $y_h=2.64(75)$ for the CNNs, and $y_{t}=1.67(9)$ and $y_h=2.21(73)$ for the FCNNs. In Figs.~\ref{fig:Meta} (e) and (f), and Table~\ref{tab:Meta_Tri}, we use the CNN and FCNN sets that yield  estimates close to known values of $y_t$ and $y_h$.

In the vicinity of the first-order boundary for the Ising model~\eqref{eq:Isingham}, we obtain the $P_{L}(T=2.1,h)$ in Fig.~\ref{fig:Ising_FO}, by averaging over all the $10$ CNNs, for each $L$ [see the discussion in Appendix~\ref{app:NNAT}].

\subsection{\label{app:Class_acc}Relation between system size, two-scale factor, reduced temperature and classification accuracy}

In Eq.~\eqref{eq:class} we had said that, if $N^{\alpha}_{L}$ is the number of spin configurations that are correctly classified out of a total $N$ configurations for a system of linear size $L$, say below $T^{\alpha}_{c}$ and in the temperature range $[T^{\alpha}_{0},T^{\alpha}_{c}]$, then 
\begin{eqnarray}
\frac{(N-N^{\alpha}_{L})}{N}\propto\frac{1}{f_{t}^{\alpha}|t^{\alpha}_{0}|L^{y_t}}\,,
\label{eq:classA}
\end{eqnarray}  
where $t^{\alpha}_{0}\equiv\frac{T_{0}-T^{\alpha}_{c}}{T_{c}}$. We derive this relation now.

Let $n_s$ be the number of spin configurations (henceforth, snapshots) that we use in our testing data set for reduced temperatures that lie between $t$ and $t+dt$, where $t$=$\frac{T-T_{c}}{T_c}$. Out of these $n_s$ snapshots, let $n_{L}(t)$ be the number of snapshots that are correctly classified (e.g., as lying below $T_c$), for a system with linear size $L$. By definition, 
\begin{eqnarray}
n_{L}(t) = \mathcal{F}_{L}(t)n_s \,,   
\end{eqnarray}
where $\mathcal{F}_{L}(t)=\int\,d\boldsymbol{\sigma}\,\mathfrak{P}_{L}(\boldsymbol{\sigma},t,0)H(\hat{P}_L(\boldsymbol{\sigma})-0.5)$, the Heaviside function $H(x)=0,\,\text{if}\,x<0,\,\text{and} \,H(x)=1,\text{if}\,x\geq0$, and we use $0.5$ as the threshold for classifying the spin configurations. Arguments similar to those used in Eqs.~\eqref{eq:PL2}, ~\eqref{eq:PL3}, and ~\eqref{eq:PL4}, give the same FSS form for $\mathcal{F}_{L}(t)$ as that for $\mathcal{P}_{L}(t,0)$.

If $n$ is the number of sets of such ($n_s$) snapshots, in the intervals $[t_i, t_i+dt]$, where $i \in [1,2, \ldots, n]$, and total number of snapshots is $N=n_s*n$, then the total number of snapshots correctly classified (e.g., as lying below $T_c$) is
\vspace{-0.1cm}
\begin{eqnarray}
N_{L} = \sum_{i=1}^{n} n_{L}(t_i)=\sum_{i=1}^{n} \mathcal{F}_{L}(t_i)\frac{N}{n}\,.
\label{eq:Psum}
\end{eqnarray}
\vspace{-0.1cm}
Using $\frac{T_c-T_0}{T_c}=ndt=|t_0|$ in equation~\ref{eq:Psum} we get
\vspace{-0.1cm}
\begin{eqnarray}
\frac{N_{L}}{N} &=& \frac{1}{|t_0|}\sum_{i=1}^{n}\mathcal{F}_{L}(t_i)dt\,, {\rm{and,\; if}}\;\; n \to \infty\,, \nonumber\\
&=&\frac{1}{|t_0|}\int_{t_0}^{0} \mathcal{F}_{L}(t)dt\, \nonumber\\
&=& \frac{1}{|t_0|L^{y_t}}\int_{x_{L}^{0}}^{0} \mathcal{F}^{*}_{L}(x)dx\,,
\label{eq:PInt}
\end{eqnarray}
\vspace{-0.1cm}
where we have used $\mathcal{F}_{L}(t)\equiv\mathcal{F}^{*}_{L}(tL^{y_{t}})=\mathcal{F}^{*}_{L}(x)$ and changed the variable of integration from $t$ to $x=tL^{y_t}$.
From Eq.~\eqref{eq:PInt} and using
\begin{eqnarray}
\int_{x_{L}^o}^{0} \mathcal{F}^{*}_{L}(x)dx&=&\int_{x_L^0}^{X^0} \mathcal{F}^{*}_{L}(x)dx+\int_{X^0}^{0} \mathcal{F}^{*}_{L}(x)dx\nonumber\\
&\approx& X^0 - x^0_{L} +\int_{X^0}^{0} \mathcal{F}^{*}_{L}(x)dx\nonumber\\
&=& X^0 - t_0 L^{y_t} + \int_{X^0}^{0} \mathcal{F}^{*}_{L}(x)dx\nonumber\\
&=& a + |t_0| L^{y_t}\,,
\label{eq:simp}
\end{eqnarray}
\vspace{-0.1cm}
where $\mathcal{F}^{*}(x)$ is asymptotically close to $1\,\;\forall x_{L}^{0}<X^{0}$, we get
\begin{eqnarray}
\frac{N_L}{N}\approx1 + \frac{a}{|t_0|L^{y_t}}\implies\frac{(N-N_L)}{N}\approx\frac{A}{|t_0|L^{y_t}}\,,
\label{eq:classF}
\end{eqnarray}  
\vspace{-0.1cm}
where $A=-a$.
If we now consider $\mathcal{F}^{*\alpha}_{L}(x)$, for the Ising ($\alpha=I$), Blume-Capel ($\alpha=BC$), and Metamagnet ($\alpha=M$) models~\eqref{eq:Isingham}, \eqref{eq:IsingBC}, and \eqref{eq:IsingMeta}, we must
include the scale factors $f_t^{\alpha}$ [see Eq.~\eqref{eq:fBCfM}], and use $\mathcal{F}^{*\alpha}_{L}(f_t^{\alpha}x)\equiv\mathcal{F}^{*}_{L}(x)$ to obtain 
\vspace{-0.1cm}
\begin{eqnarray}
a^{\alpha}=X^{0,\alpha}+\int_{X^{0,\alpha}}^{0} \mathcal{F}_{L}^{*\alpha}(x')dx'\nonumber\\
=\frac{X^{0}}{f_t^{\alpha}}+\frac{\int_{X^{0}}^{0} \mathcal{F}^{*}_{L}(x)dx}{f_t^{\alpha}} = \frac{a}{f_t^{\alpha}}\,\nonumber\\
\label{eq:af}
\end{eqnarray}
Then the relation for the number of correctly classified snapshots for $t_0^{\alpha}\leq t \leq0$, $N_{L}^{\alpha}$ is 
\begin{eqnarray}
\frac{(N-N^{\alpha}_{L})}{N}&=&\frac{A}{f_t^{\alpha}|t^{\alpha}_0|L^{y_t}}\,. 
\label{eq:class_App}
\end{eqnarray}   
We verify this relation in Fig.~\ref{fig:log_log} (b), where we plot $\frac{(N-N^{\alpha}_{L})}{N}$ vs $f_t^{\alpha}|t^{\alpha}_{0}|L^{y_t}$ for the Ising ($I$)~\eqref{eq:Isingham}, Blume-Capel ($BC$)~\eqref{eq:IsingBC} and Ising-metamagnet ($M$)~\eqref{eq:IsingMeta} models, for systems with linear sizes $L=12,\,16,\,20$ and $30$ [in Fig.~\ref{fig:log_log} (b), we use the temperature range $[T_c-1,T_c]$ for the Ising~\eqref{eq:Isingham} and Blume-Capel ~\eqref{eq:IsingBC} models, 
and $[T_c-1.5,T_c]$ for the Ising-metamagnet model~\eqref{eq:IsingMeta}].

\bibliography{main}

\end{document}